\documentstyle[11pt]{article}
\newcommand{\rdot}{\dot{r}}
\newcommand{\vdot}{\dot{v}}

\newcommand{\rds}{\dot{r}^2}
\newcommand{\Rds}{\dot{R}^2}
\newcommand{\ddv}{\left( \frac{\partial}{\partial v}\right)^{\! a} }
\newcommand{\ddr}{\left( \frac{\partial}{\partial r}\right)^{\! a} }

\newcommand{\eps}{\varepsilon}
\newcommand{\e}{\varepsilon}
\newcommand{\ord}{{\mathcal{O}}}

\newcommand{\om}{\omega}

\begin{document}


\bigskip\bigskip\bigskip
\centerline{\Large \bf Overcharging a Black Hole and  Cosmic Censorship}
\bigskip\bigskip
\centerline{\bf Veronika E. Hubeny}
\medskip
\centerline{Physics Department}
\centerline{University of California}
\centerline{Santa Barbara, CA. 93106}
\medskip
\centerline{email: veronika@cosmic.physics.ucsb.edu}

\begin{abstract}

We show that, contrary to
a widespread belief, one can overcharge
a near extremal Reissner-Nordstr\"{o}m black hole by throwing in 
a charged particle, as long as the backreaction effects may
be considered negligible.
Furthermore, we find that we can make the 
particle's classical radius, mass, and charge, as well as
the relative size of the backreaction terms 
arbitrarily small, by adjusting the parameters
corresponding to the particle appropriately.
This suggests that the question of cosmic censorship is still
not wholly resolved even in this simple scenario.
We contrast this with attempting to overcharge a
black hole with a charged imploding shell, where we find that
cosmic censorship is upheld.  We also briefly comment on 
a number of possible extensions.

\end{abstract}

\pagebreak

\section{Introduction}

The question of cosmic censorship remains one of the 
most important open questions in classical general relativity.
Since Penrose proposed the idea in 1969~\cite{penr},
numerous papers have been produced addressing the question
of whether the cosmic censorship conjecture holds or not.
We shall not attempt to review those developments here;
for some of the recent reviews and lists of references, 
see e.g.~\cite{w2},~\cite{c},~\cite{sin}.
Rather, we will consider a specific attempt to violate cosmic
censorship.  To that end, we will first give a brief discussion
of what the conjecture 
implies in this context and therefore
what might constitute providing a
counter-example to it. 

The basic idea of cosmic censorship is roughly as follows:
Starting from regular, generic initial conditions, and evolving
the system using classical general relativity, 
we cannot obtain a singularity which can be observed from ``far away''
(or more precisely, from the future null infinity).\footnote{
This is referred to as the ``weak'' cosmic censorship. 
The ``strong'' version drops the ``far away'' proviso, namely,
it asserts that we cannot evolve to a timelike singularity.
In the following, we will consider just the ``weak'' version.}
On the other hand, we know from the singularity theorems 
that singularities are a rather generic occurrence in general 
relativity.  For instance, we can obtain a singularity from regular
and generic initial data (with ``reasonable'' properties, such as the
energy condition) just by concentrating a sufficient amount of matter
into a given compact region of space.
The statement of cosmic censorship then implies that one cannot
destroy a black hole (i.e.\ its horizon) just by throwing in
more ``reasonable'' matter.
We will attempt to test the validity of this statement by
considering a particular example.

In his seminal paper, Wald~\cite{w1} attempted to 
destroy a Kerr-Newman black hole by 
throwing in particles which ``supersaturate'' extremality
(i.e. which, when captured by the black hole, would
violate the bound $M^2 \geq Q^2 + a^2$, where 
$M$ is the mass, $Q$ the charge, and $a$ the
angular momentum per unit mass of the resulting
configuration).
The Kerr-Newman metric with such a bound violated 
does not possess an event horizon, and therefore 
describes a naked singularity.
Wald presented gedanken experiments 
of an extremal black
hole capturing particles with high charge
and angular momentum to mass ratio.
However, he found that precisely those particles
which would exceed extremality will not be 
captured by the black hole.

In the spirit of~\cite{w1}, we present a mechanism
whereby a black hole does ``capture'' a particle such that the
parameters of the final configuration  correspond to
a naked singularity.
The crucial difference between our set-up and that of~\cite{w1}
is that we are starting with a black hole arbitrarily close to, 
but not exactly at, extremality.

While our calculation 
of the ``test'' process (i.e. neglecting backreaction effects)
unequivocally indicates that 
the particle falls into the singularity and overcharges the 
black hole, the implications about cosmic censorship
cannot be easily assessed without 
first estimating the backreaction effects.
We attempt to do so in this paper.  Although we  find these
to be small, 
a definite conclusion with regards to cosmic
censorship is not yet at hand.

We contrast this situation with that of attempting to overcharge
a black hole with an imploding charged shell.   
In such a case of exact spherical symmetry, we can determine the
equation of motion exactly--i.e.\ we do not need to consider 
further backreaction effects or be confined to just ``test''
processes.  However, we find that in this case the self-repulsion
of the shell prevents it from collapsing past the horizon 
of the black hole.

The outline of this paper is as follows:
In section \ref{bas} we describe the basic idea for
overcharging a black hole.
In section \ref{mech}, we discuss the actual mechanism, 
and provide a numerical example.
Section \ref{shell} is devoted to  the discussion of the simpler
case of an imploding charged shell.
Various extensions and other examples of potential cosmic
censorship violation
are presented in  section \ref{ext}, and we discuss and 
summarize our results in section \ref{concl}.

\section{Basic Idea}
\label{bas}

The basic general idea for ``overcharging''
a  black hole is the following:
We start with a near-extremal Reissner-Nordstr\"{o}m black hole with
charge $Q$ and mass $M$.  
The goal is to send in (radially) a charged particle 
with charge $q$, rest mass $m$, and energy $E$,
such that $m < E < q \ll Q < M$, satisfying the following 
conditions:

\begin{enumerate}
\item{The particle falls past the radial coordinate, $r_{+}$, corresponding
to the initial horizon of the black hole.}
\item{The final configuration exceeds extremality, $Q+q > M+E$.}
\item{The particle may be treated as a test particle; 
that is, the ``backreaction'' effects are negligible.}
\end{enumerate}
(Whereas conditions 1.\ and 2.\ would in principle suffice
to overcharge a black hole,
condition 3.\
renders our calculations  meaningful and feasible.)

Then, after the completion of such a process, 
the solution should approach a spacetime
described by Reissner-Nordstr\"{o}m
metric with 
$M_{{\rm final}} \leq M+E < Q+q \equiv Q_{{\rm final}}$,
which corresponds to a naked singularity
rather than a black hole.\footnote{
We write $M_{{\rm final}} \leq M+E$ in order to include 
the possibility that some of the mass is carried away by
gravitational radiation before the system settles down to its 
final state.}
In such a case, our
process seems to ``destroy the horizon of a black hole'',
in violation of cosmic censorship.

It has long been known (\cite{w1},\cite{co},\cite{semiz})
that one cannot destroy a black hole using this process in a 
quasistationary manner, that is, 
by allowing  the black hole mass and charge to change
just by infinitesimal amounts.
Namely, if one considers strictly test
processes, to overcharge a black hole, one must first reach
extremality.  But once extremality is reached, any further
changes would at best maintain it.
The crucial point here is that in our process,
the change in black hole's parameters is finite (albeit small), 
so that we are not constrained to starting with an extremal 
black hole in order to overcharge it.

\section{The Mechanism}
\label{mech}

In this section we discuss how
well we can satisfy conditions 1.--3.
Namely, for a suitable choice of  parameters $(M,Q)$ 
corresponding to the black hole,
 we wish to find the parameters  $(m,E,q)$ describing the particle
which satisfy all three conditions.
The specification of these 5 parameters 
determines the  motion of the particle.

We proceed in two steps.  If we can find parameters which
satisfy all three conditions, then in particular,
the particle's motion will be described arbitrarily well by the 
test charge equation of motion. 
 Hence we can show that conditions 1.\ and 2.\ hold
(for a suitable choice of parameters)
 using the test particle approximation (i.e.\ {\em assuming}
condition 3.), and then attempt to  show 
the self-consistency of the approach by
arguing that the backreaction effects can be arbitrarily small.

Such calculation however still does not suffice to argue
that all three conditions are satisfied {\em simultaneously},
due to the perhaps somewhat subtle reason that 
a small change in the equation of motion can produce 
large changes in the trajectory.
In fact, we will see that even though the test approximation leads to 
conditions 1.\ and 2.\ being fulfilled {\em and}
condition 3.\ is satisfied to a desired degree of
accuracy, using a better approximation to the particle
equation of motion leaves 
the fulfilment of condition 1.\ in question.

Nevertheless, 
studying conditions 1.\ and 2.\ separately may  be of some
interest on its own.  Although as pointed out in the Introduction,
we cannot make claims about real cosmic censorship violation
without taking backreaction into account, the ``test'' analysis
will serve to provide a comparison with earlier works such 
as~\cite{w1}, and will be used as a springboard for the
analysis of the backreaction effects.

\subsection{Motion of a test charge}
\label{test}

We consider the motion of a test particle
of charge $q$, mass $m$, and four-velocity  $u^{a}$,
in a fixed background Reissner-Nordstr\"{o}m
spacetime 
with parameters $M$ and  $Q$, 
and an electrostatic potential $A_{a}$.
In the ingoing Eddington coordinates, $(v,r,\theta,\varphi)$,
the Reissner--Nordstr\"{o}m metric is
\begin{equation} 
\label{ds}
ds^2 = - g(r) \, dv^2 + 2 \, dv \, dr + r^2 \, d\Omega^2
\end{equation} 
where 
$ d\Omega^2 = d\theta^2 + \sin^2 \! \theta \, d\varphi^2 $ 
and
$g(r) \equiv 1 - \frac{2M}{r} + \frac{Q^2}{r^2}$.
For $Q \leq M$, the metric describes a black hole
with charge  $Q$ and mass $M$.
In this case we can write
$g(r) = \frac{(r - r_{+})(r - r_{-})}{r^2}$
with the event and Cauchy horizon 
given by 
$r_{\pm} = M \pm \sqrt{M^2 - Q^2}$.
Hence $r_{-} \leq Q \leq M \leq r_{+}$.
Extremality corresponds to $Q = M$,
whereas for $Q > M$, we have a naked singularity.
The electrostatic potential $A_{a}$ has the only 
nonvanishing component
$A_{v}(r) = - \frac{Q}{r} $.

The equation of motion for a test charge is given by
\begin{equation} 
\label{eom}
u^{a} \, \nabla_{\! a}\, u^{b} = \frac{q}{m}\, F^{b}_{\ c}\,  u^{c}
\end{equation} 
with $F_{ab} \equiv 2 \nabla_{\! [a} A_{b]}$.
For radial trajectories, we can write 
$u^{a} = \vdot \ddv  + \rdot \ddr $
where $\ \dot{ } \, \equiv \frac{d}{d\tau}$ and $\tau$ is the
affine parameter along the worldline.
Since $\ddv$ is a Killing field, the particle's ``energy'',
$E \equiv - \ddv \, (m u_{a} + q A_{a})
= m \, \left[ g(r) \, \vdot - \rdot \right] - q A_{v}$
is a constant of motion
 along the particle's worldline.
The particle follows a timelike trajectory, so that
$-1 = u^{a} u_{a} = 
- g(r) \, \vdot^2 + 2 \vdot \rdot =
- \frac{1}{m} (E + q A_{v}) \vdot +  \vdot \rdot$.
Eliminating $\vdot$ from
these expressions and solving for $\rdot$,
we obtain the radial equation of motion in a particularly
useful form:
\begin{equation} 
\label{reom}
\rds =  
\frac{1}{m^2} \left[ E + q A_{v}(r) \right]^2 - g(r).
\end{equation} 

Since $\rdot = 0$ corresponds to a turning point,
the particle ``falls in'' if $\rds$ never vanishes 
outside the horizon,
\begin{equation} 
\label{rdscond}
\rds > 0 \ \ \ \ \ \forall \  r \geq  r_{+}.
\end{equation} 
Thus, if we find suitable parameters 
$m$, $E$, and $q$, 
such that eq.(\ref{rdscond}) is satisfied, 
then condition 1.\ will hold.
Since 
$g(r)$ is positive outside the horizon,
eqs.\ (\ref{reom}) and (\ref{rdscond}) imply that 
$\left[ E + q A_{v}(r) \right]^2 > m^2 \, g(r) > 0 
\ \ \forall \ r \geq r_{+}$.
This imposes the necessary condition on the energy:
$E > -q \, A_{v}(r) =  \frac{qQ}{r} $.
(Note that this relation also follows from 
the above definition of the energy,
since $g(r) \, \vdot - \rdot > 0$ initially.)
We can simplify this condition by noting that 
$-q \, A_{v}(r)$ attains its maximal value at $r = r_{+}$.
Replacing $r$ with $r_{+}$ above,
 we have
\begin{equation} 
\label{emin}
E > \frac{q Q}{r_{+}}.
\end{equation} 
Condition 1.\ thus provides a lower bound on $E$.
On the other hand, condition 2.\ provides an upper bound:
\begin{equation} 
\label{emax}
E < Q + q - M.
\end{equation} 
Thus, in order to find $E$ satisfying both conditions,
$q$ must satisfy the inequality
$\frac{q Q}{r_{+}}
< Q + q - M$,
or
\begin{equation} 
\label{qpick}
q > r_{+}  \left( \frac{M-Q}{r_{+}-Q} \right) 
= \frac{r_{+}-Q}{2}.
\end{equation} 

Note that for a non-extremal black hole, we can always find 
sufficiently large $q$ such that there exists a finite range of 
allowed energies $E$. 
Furthermore, the allowed range of admissible $E$'s grows linearly
with $q$.
On the other hand, if we start with an extremal black hole, 
$Q=M= r_{+}$, then we obtain $E>q$ from eq.(\ref{emin}), and
$E<q$ from eq.(\ref{emax}), which means that conditions
1.\ and 2.\ cannot be satisfied simultaneously---i.e.\ we
cannot overcharge an extremal black hole.

The above choice of $q$ and $E$ automatically ensures
that condition 2.\ is satisfied.  To address condition 1., 
we must return to eq.(\ref{reom}).
Since $g(r)$ is bounded and 
$\left[ E + q A_{v}(r) \right]^2 > 0 \ \  \forall \  r \geq r_{+} $,
we can always find $m$ such that $\rds > 0$.
In particular, we must pick 
$m < (E - qQ/r)/\sqrt{g(r)}
 \ \  \forall \   r \geq  r_{+} $.
Analytically, the RHS  attains its minimum at
\begin{equation} 
\label{mmin}
\stackrel{-}{r}_{m} = 
 Q \left( \frac{Mq-QE}{Qq-ME} \right) > r_{+},
\end{equation} 
where the inequality follows from eq.(\ref{emin}).
Upon substitution, we find that $m$ must satisfy the bound
\begin{equation} 
\label{mpick}
 m < Q \sqrt{\frac{-E^2+2 \frac{M}{Q} Eq -q^2}{M^2-Q^2}}.
\end{equation} 
Note that the positivity of the radical follows automatically
from the above conditions:
\begin{eqnarray} 
-E^2+2 \frac{M}{Q} Eq -q^2 
= 2Eq \left( \frac{M}{Q} -1 \right) - (q-E)^2 \nonumber \\
> (M-Q)^2 \left[ 2 r_{+} \frac{M-Q}{(r_{+}-Q)^2} - 1 \right]
=0.
\end{eqnarray}
This choice of $m$  ensures that eq.(\ref{rdscond}) holds.

To summarize, given the background spacetime with parameters $(Q,M)$,
we can satisfy conditions 1.\ and 2.\
by choosing the parameters $m$, $E$, and $q$
as follows:
\begin{enumerate}
\item{Pick $q$ which satisfies eq.(\ref{qpick})
with the given values of $Q$ and $M$.}
\item{Select $E$ satisfying eqs.\ (\ref{emin}) and (\ref{emax}) with
the chosen $q$, $Q$, and $M$.}
\item{Finally, choose $m$ according to eq.(\ref{mpick}) 
using the other parameters.}
\end{enumerate}
This prescription guarantees that if the particle
follows the test charge equation of motion (as dictated by
condition 3.), then it falls past the horizon and
``overcharges'' the black hole.
Also note that the parameters necessary for this process
are not too restricted: $q$ is bounded only from below,
the range of allowed values of $E$ grows linearly with $q$, 
and $m$ is bounded only from above.
Although condition 3.\ will restrict these parameters further,
their ``generic'' nature will be preserved.

It may be of interest to consider what happens to the particle
after it crosses the horizon.\footnote{
Of course, as will be discussed below,  we expect the backreaction
effects to take over sufficiently close to the singularity,
 so the {\em actual} behavior may be very
different from that predicted by the test equation of motion.}
In particular, does it fall into the singularity, or does it 
``bounce'' and reemerge from the ``horizon'' (possibly into another
asymptotically flat region of the spacetime)?
In order for the particle to fall to $r=0$, it must satisfy
$\rdot(r) < 0 \ \ \forall r$, 
and in particular, 
$\rds(r_{m}) > 0$ with $r_{m}$ denoting the minimum of $\rds(r)$.
In the above construction we only required
$\rds(r) > 0 \ \ \forall r \geq r_{+}$.
However, eq.(\ref{emin}) implies that $\left( E- \frac{qQ}{r} \right)$
could only vanish at radius $r_{0} > Q > r_{-}$.
For  $r_{0} < r_{+}$ we then have $g(r_{0}) < 0$, i.e.\ $\rds(r_{0}) > 0$,
whereas  $r_{0} >  r_{+}$ corresponds to the case discussed above.
In either situation, eq.(\ref{mpick}) then guarantees that 
$\rds > 0 \ \ \forall r$, which means that once the particle passes 
the horizon, it also falls into the singularity.
To see this more clearly,
it will be convenient to rewrite
eq.(\ref{reom}) as
\begin{equation} 
\label{reom1}
\rds =  
 \left( \frac{E^2}{m^2} - 1 \right)
- \frac{2M}{r} \left( \frac{Q}{M} \frac{Eq}{m^2} - 1 \right)
+ \frac{Q^2}{r^2} \left( \frac{q^2}{m^2} - 1 \right).
\end{equation} 
The minimum is attained at
\begin{equation} 
\label{reomin}
r_{m} =  
 Q \left( \frac{q^2}{m^2} - 1 \right)
 \left( \frac{Eq}{m^2} -  \frac{M}{Q} \right)^{-1},
\end{equation} 
with the corresponding value 
\begin{equation} 
\label{rdsmin}
\rds(r_{m}) =  
 \left( \frac{q^2}{m^2} - 1 \right)^{-1} \left\{
- \left( \frac{q-E}{m} \right)^{2}
+ 2 \left( \frac{M}{Q} - 1 \right) \left[
\frac{Eq}{m^2} -  \frac{1}{2} \left( \frac{M}{Q} + 1 \right)
\right] \right\}.
\end{equation} 
Eq.(\ref{mpick}) then dictates that $\rds(r_{m}) > 0$.

We also note in passing that the conclusion reached above 
about extremal black holes becomes even more transparent in
this formulation:  for $Q=M$, $r_{m} > r_{+}$, and 
$\rds(r_{m}) < 0$ since the second term on RHS of 
eq.(\ref{rdsmin}) vanishes.

Hence, the test charge equation of motion dictates that 
with a suitable choice of parameters, the 
particle not only passes the initial horizon of the black hole, 
but also that it crashes into the singularity at $r=0$.
If this were indeed an accurate description of the actual process,
then it would be difficult to see how the cosmic censorship could
survive.

\subsection{Backreaction effects}
\label{backreaction}

The proposed mechanism for overcharging a black hole can
only be valid if we can trust our conclusions, in particular,
if the use of the test charge equation of motion for our
 particle is justified.  
It has been shown (e.g.\ by \cite{br}, \cite{n}) that in some instances,
the ``backreation'' effects do become important, so that the 
test particle calculation can no longer be trusted.
Thus, to check whether such effects  arise in our case, 
we have to examine the backreaction.

In the following, 
we  first present rough plausibility arguments for 
the negligibility of the backreaction effects,
followed by a more systematic analysis.
Finding that condition 3.\ may be fulfilled to an arbitrary
degree of accuracy, we  then reexamine condition 1.\
using a better approximation to the equation of motion.
Since a full analysis of the corrected solution in the general
situation is beyond the scope of this work, 
we present a numerical example in section \ref{numex}.

There are several different meanings of ``backreaction'',
depending on the context.
First, the presence of the particle modifies the spacetime metric, 
so that the approximation of the particle moving on a fixed
background spacetime need no longer be valid. 
As a rough indicator of the  magnitude of this effect,
we may compare the relative 
sizes of the background electromagnetic 
stress-energy tensor with the electromagnetic stress tensor
of the particle;  if the latter is much smaller
than the former, the approximation of the particle
moving on a fixed background spacetime should be a
good one.  
 
Assuming the particle may be thought of as a uniformly
charged ball of dust (in particular, that the internal stresses
are not significant\footnote{
We believe that this is a physically reasonable 
assumption, since we can first consider a particle satisfying
these conditions, then take the black hole mass to be
much bigger that the particle mass or charge, and then set the
black hole charge $Q$ to be near enough to
extremality such that our conditions are satisfied.
}), 
the particle's stress tensor is maximal at its
 ``surface'' at radius $r_{q}$, where 
$|T_{\rm q}^{ab}| \sim F_{\rm q}^2 \sim 
\left( \frac{q}{r_{q}^2} \right)^2 $.
In contrast, the electromagnetic stress tensor of the
black hole at radius $r$ is
$|T_{\rm BH}^{ab}| \sim F_{\rm BH}^2 \sim 
\left( \frac{Q}{r^2} \right)^2  \sim \frac{1}{M^2}$.
(The last relation holds only in the vicinity of the black hole,
but that is the regime we are mainly interested in.)
Thus, to satisfy $|T_{\rm q}^{ab}| \ll |T_{\rm BH}^{ab}|$,
we need to allow 
$r_{q}^2 \gg  q M$.
At the same time, we have the obvious requirement that
$r_{q} \ll  M$, i.e.\ the particle must be smaller than
the black hole in order for us to talk meaningfully of it 
falling into the black hole.  Also, 
 to follow the same equation of motion,
its constituent parts must
be subjected to the same background curvature.
As we will show explicitly below, we can satisfy
both conditions for sufficiently small $q$.

Since there exists a consistent choice of $r_{q}$ satisfying
$\sqrt{q M} \ll r_{q} \ll M$, and since the equation of motion
is independent of this choice of $r_{q}$,  we will henceforth
treat the particle as 
moving on a fixed background spacetime.
Strictly speaking, 
this argument  may not be sufficient to justify neglecting this
form of backreaction 
(the reason will become more clear at the end of this section);
however, further analysis is beyond the scope of this paper.

In the following, we will concentrate on a  somewhat weaker
form of ``backreaction'',
involving only a modification of the particle's equation
of motion on a fixed background spacetime.
(Hence the treatment of this latter form of backreaction is contingent
upon the negligibility of the former effects.)

To this end, we may use the results 
obtained recently by Quinn \& Wald~\cite{qw}
(and previously by e.g.\ \cite{dwb}, later corrected by \cite{hobbs}),
who calculate 
the corrections to the test particle equation of motion due
to its ``self field'' effects, on a fixed spacetime.
Using $\frac{q^2}{m}$ as a small expansion parameter,
\cite{qw} derive the corrected equation of motion
for a small spherical charged body to the leading order:
\begin{eqnarray} 
\label{acorr}
& a^{a} \equiv u^{b} \, \nabla_{\! b}\, u^{a}   \nonumber \\
& = \frac{q}{m} \, F^{ab} \,  u_{b}
+ \frac{2}{3} \, \frac{q^2}{m} \, \left(
\frac{q}{m} \, u^{c} \, \nabla_{\! c} \, F^{ab} \, u_{b} 
+ \frac{q^2}{m^2} \,  F^{ab} \,  F_{bc} \, u^{c}
- \frac{q^2}{m^2} \, u^{a} \, F^{bc} \, u_{c} \,  F_{bd}
\, u^{d} \right) \nonumber \\
& + \frac{1}{3} \, \frac{q^2}{m} \, \left(
 R^{a}_{\ b} \, u^{b} +  u^{a} \, R_{bc} \, u^{b} \, 
 u^{c} \right)
+ \frac{q^2}{m} \,  u_{b} \, \int_{-\infty}^{\tau} 
\nabla^{[b}(G^{-})^{a]c} \, u_{c}(\tau') \, d\tau'
\end{eqnarray} 
Here $F^{ab}$ is the background Maxwell tensor (corresponding to 
 the electromagnetic field of the black hole,
 but not the particle's
self-field), $R^{ab}$ is the Ricci curvature tensor
for the background metric, and $(G^{-})^{ab}$ is the
retarded Green's function for the vector potential
in the Lorentz gauge.

The first term on the RHS corresponds to the test particle limit
(cf.\ eq.(\ref{eom})).
The next three terms describe the Abraham-Lorentz radiation
damping, the following two local curvature terms 
are present to preserve the conformal invariance of the
acceleration, and the last term is the ``tail term'',
which arises due to the failure of the Huygen's
principle in curved spacetime.\footnote{
There are also $\ord(m)$ corrections 
such as the gravitational radiation
corrections (which~\cite{qw} treat separately), but since
$q>m$, i.e.\ $\frac{q^2}{m} > m$, we expect these terms to be
subleading.}
Note that all the correction terms 
 contain a factor of $\frac{q^2}{m}$.

Hence we clearly desire to make $\frac{q^2}{m} \ll M$.
We may interpret this physically as making the ``classical
radius'' of the particle small compared to the size
of the black hole.
(The ``classical radius'' is defined by 
 $r_{cl} \equiv  \frac{q^2}{m}$,
corresponding to the mass $m$ of a (spherical) particle given
 entirely by its electrostatic self-energy, $\frac{q^2}{r_{cl}}$.)
The condition $\frac{q^2}{m} \ll M$ is then imposed by the
previous argument:
By the positive energy condition, we require the 
electrostatic self-energy of the particle to be
smaller than or equal to its rest mass, so that 
 $\frac{q^2}{r_{q}} < m$, or $r_{cl} < r_{q} $.
Hence to satisfy $r_{q} \ll M$, we need
$\frac{q^2}{m} = r_{cl} \ll M$.

Although we will show explicitly how
to make $\frac{q^2}{Mm}$ arbitrarily small,
 we can see that a consistent way to choose $q$ and $m$ 
should exist, based on the following heuristic argument:
For the test particle approximation, it is clear
that we need to take $q \ll Q$.  This in turn means
that we must start near extremality, so that
the range of allowed $E$'s is small. 
As is apparent from eqs.\ (\ref{emin}) and  (\ref{emax}),
as $Q \rightarrow M$, any allowed $E \rightarrow q$.
Setting  $E \sim q$ in eq.(\ref{mpick}), we see that conditions
1.\ and 2.\  then only  constrain the
ratio $\frac{q}{m}$, i.e.
\begin{equation} 
\label{mpick2}
\frac{q}{m} \stackrel{\sim}{>} \frac{\sqrt{g(r)}}{1 - \frac{Q}{r}}
\ \ \forall \ \ r \geq r_{+}, \ \ 
{\rm or } \ \frac{q}{m} \stackrel{\sim}{>} \sqrt{\frac{1}{2} \left(
\frac{M}{Q}+1 \right)},
\end{equation} 
but do not impose any further constraints on $\frac{q^2}{m}$.
We thus expect that if we start with the
black hole near enough to extremality, so that we
can take $q$ to be very small, then we will satisfy
$q \, \frac{q}{m} \ll M$.
A more careful analysis below shows that this is indeed 
the case
(and a numerical example is given in section \ref{numex}).

However, although the expansion parameter is small, 
we still need to check that the actual correction terms
themselves are small.   To that end, we must consider 
eq.(\ref{acorr}) in greater detail.

We find that for radial motion in a static potential,
the third and fourth terms on the RHS of eq.(\ref{acorr}) 
cancel:
$ F^{ab} \,  F_{bc} \, u^{c}
= u^{a} \, F^{bc} \, u_{c} \,  F_{bd}
\, u^{d}$.
\\ Correspondingly, the curvature terms 
(i.e.\ fifth and sixth terms) also cancel,
\begin{equation} 
 R^{a}_{\ b} \, u^{b} +  u^{a} \, R_{bc} \, u^{b} \, 
 u^{c} 
= 2 \left( F^{ab} \,  F_{bc} \, u^{c}
-u^{a} \, F^{bc} \, u_{c} \,  F_{bd}
\, u^{d} \right) = 0.
\end{equation} 

The tail term requires a bit more analysis, since it involves
the Green's function, and therefore is a-priori nonlocal.
However, since 
it would not arise in a flat background spacetime,
 and we can set the background curvature as small as we need
by letting $M$ be large,
we expect there to be a regime in which
the leading order effect is given only in terms of the local physics.
Indeed, as~\cite{qw} point out, for sufficiently 
small spacetime curvature and slow motion of the body,
``one would expect the `tail term' to become effectively
local, since the contributions to the `tail term' arising from
portions of the orbit distant from the present position of the
particle should become negligible.''
We will thus {\em assume} that the full tail term is roughly bounded by 
its local part; specifically we assume that the full integral is
not much larger than the largest value of the integrand (multiplied
by a unit $\Delta \tau$ to make the units comparable).

The local expression for the Green's function 
in curved spacetime may be obtained
from e.g.\ \cite{dwb}. 
In the current case of vanishing Ricci scalar, we may write
this local contribution as
$\nabla^{[b}(G^{-})^{a]c} = \nabla^{[b}R^{a]c}$.
In terms of the Maxwell tensor and the components of the 4-velocity,
we can then write this, with the actual factors appearing 
in the equation of motion as
\begin{equation} 
\label{tail}
 - \frac{q^2}{m} \,  u_{b} \, u_{c} \, \nabla^{[a}R^{b]c}
= \left( \frac{q}{m} \, F^{ab} \,  u_{b} \right) \, \left[
-q f' (g \vdot -\rdot) \right],
\end{equation} 
 where we have employed the notation
$f(r) \equiv F^{vr} = \frac{Q}{r^2}$, so
$f'(r) = \frac{df}{dr} = - \frac{2Q}{r^3}$.
Since $(g \vdot -\rdot) = (E-\frac{qQ}{r})/m$, which is small 
near the horizon and $\ord(1)$ further away, and
$f' \sim \ord(1/M)$ at $r \sim r_{+}$ and falls off
as $r^{-3}$, the whole term should
have at most a contribution 
$\sim \frac{q}{M} $
near the horizon, which is indeed tiny.

Finally, let us consider the second term, which we will refer to as
the ``radiation damping'' term.
Based on the above discussion, we expect this to provide the
dominant correction to the test particle equation of motion.
We can rewrite 
the radiation damping in the form  
\begin{equation} 
\label{damp}
\frac{2}{3} \, \frac{q^3}{m^2} \, 
u^{c} \, \nabla_{\! c} \, (F^{ab}) \, u_{b}= 
\left( \frac{q}{m} \, F^{ab} \,  u_{b} \right) \,
\left[\frac{2}{3} \, \frac{q^2}{m} \,
\left( \rdot \, \frac{f'}{f} + g' \, \vdot \right) \right]
\end{equation} 
where $g'(r) = \frac{dg}{dr} = \frac{2M}{r^2} - \frac{2Q^2}{r^3}$,
and we may conveniently express 
$\vdot = \frac{1}{g} (\sqrt{\rds + g} - \sqrt{\rds})$.
 Again,
we expect that the factor 
$\left( \rdot \, \frac{f'}{f} + g' \, \vdot \right)
\sim  \ord(1/M) $ 
near the horizon and falls off with $r$. 
This will be explicitly confirmed  momentarily.
The relative contribution of the radiation damping term to
the equation of motion will then  scale as
$\sim \frac{q^2}{mM}$, 
which may be arbitrarily small,  as we argued above.

To summarize, we may write the corrected 
equation of motion as
\begin{equation} 
\label{eomcorr}
u^{b} \, \nabla_{\! b}\, u^{a} =  
  \frac{q}{m} \, F^{ab} \,  u_{b} \, \left(
1 + h(r)  + \mbox{``tail''} \right)
\end{equation} 
where
$h(r) \equiv \frac{2}{3} \, \frac{q^2}{m} \, \left(
\rdot \, \frac{f'}{f} + g' \, \vdot \right)$,
and the local (dominant) contribution of the  ``tail''
is given by eq.(\ref{tail}).
We have given heuristic arguments for why we expect these
two backreaction terms to be subleading, and in fact, 
arbitrarily small.  This suggests that 
we may indeed select the parameters of the particle
such that they satisfy condition 3.

We now proceed to a more
 systematic way of showing
explicitly how,
in the regime where the test approximation fulfills condions 
1.\ and 2.,
 condition 3.\ may also be satisfied to any desired
degree of accuracy.
Namely, we use a linearized expansion around the test particle
limit.
We may normalize $M \equiv 1$ and set $Q \equiv 1-2\e^2$,
 treating $\e$ as a small parameter.\footnote{
In fact, $\eps \ll 1$ is required by conditions 2.\ and 3.\ and the
condition that we are starting with a near-extremal black hole, 
$Q<M$.}
Following the previous approach we can then reexpress 
all the other quantities in terms of $\e$.
The lower bound on $q$ imposed by eq.(\ref{qpick}) becomes\footnote{
Although we are quoting the relevant quantities only up to the 
subleading order, it is necessary to keep higher orders for the
intermediate calculations.}
$q>\e + \e^2 - \ord(\e^{3})$.
Thus we let 
$q \equiv a \e$ with $a > 1+\e$.
This allows us to find the corresponding bounds on $E$, using
eq.(\ref{emin}), which yields $ E > a \e -2 a \e^2 + \ord(\e^{3})$,
 and eq.(\ref{emax}), which yields $E <  a \e -2 \e^2$.  
Correspondingly, we let $E \equiv a \e - 2 b \e^2$ 
with $1 < b < a(1+\e)$.
Finally, from eq.(\ref{mpick}) we obtain the upper bound on $m$:
$ m < \sqrt{a^2-b^2} \e - \frac{ab}{\sqrt{a^2-b^2}} \e^2 + \ord(\e^{3})$,
which allows us to set
$m  \equiv c \, \e$ with 
$c <  \sqrt{a^2-b^2}\, (1 + \frac{ab}{a^2-b^2} \e)$.

In summary, to satisfy conditions 1.\ and 2., we may
select the parameters of the particle (to the leading order,
i.e.\ for $0 < \e \ll 1$) as follows:
\begin{eqnarray}
\label{Meps}
 M & \equiv &  1   \\ 
\label{Qeps}
 Q  & \equiv  & 1 - 2\e^2   \\
\label{qeps}
 q  &  \equiv  & a \e  
  \ \ \ \ \ \ \ \ \ \ \  \ \ \ \ 
 {\rm with } \  \   a > 1  \\
\label{Eeps}
 E  & \equiv  & a \e - 2 b \e^2  
    \ \ \ \ \ \  
 {\rm with } \  \   1 < b < a  \\
\label{meps}
 m  &  \equiv  & c \e  
    \ \ \ \ \ \ \ \ \ \ \ \ \ \ \ \ 
 {\rm with } \  \   c <  \sqrt{a^2-b^2}
\end{eqnarray}

We are now in the position to discuss condition 3.\ more explicitly.
First, we may check that the effect of the particle on the spacetime
metric is arbitrarily small near the horizon.  In particular, the condition
$\sqrt{qM} \ll r_{q} \ll M$ may be achieved
by picking, for instance, $r_{q} \sim \e^{1/4}$.  Then the particle
is small, and its stress energy is also small: 
$T_{\rm q}/T_{\rm BH} \sim \e$.
More importantly, the classical radius of the particle
is small: 
\begin{equation} 
\frac{q^2}{m} = \frac{a^2}{c} \, \e.
\end{equation} 
This means that the expansion used in eq.(\ref{acorr}) is a valid one.

From eqs.\ (\ref{damp}) and (\ref{tail})
we can also estimate the relative sizes of the
radiation damping term and the bound on the tail term.
The radiation damping term becomes:
$$ h(r) = \frac{2}{3} \, \frac{q^2}{m} \, \left(
   \rdot(r) \, \frac{f'(r)}{f(r)} + g'(r) \, \vdot(r) \right)$$
\begin{equation} 
 = \frac{4a^2}{3c^2r} \left\{ \left[ \sqrt{a^2-c^2}+
   \frac{a-2 \sqrt{a^2-c^2} }{r} \right] \e +
   \frac{2b}{\sqrt{a^2-c^2}} \left[ -a +
   \frac{a-\sqrt{a^2-c^2}}{r-1} \right] \e^2   \right\}.
\end{equation} 
Although one may worry about the divergence of the term 
$\propto \frac{1}{r-1}$
in the subleading order as $r \rightarrow 1$, one can check that in fact,
at the horizon, $r-1 \approx 2\e$, so
 $h$ is still small:
$h(r_{+})= \ord(\e)$.
Furthermore, 
the local contribution from the ``tail'' term is
\begin{equation} 
-q f' (g \vdot - \rdot) = \frac{2a^2}{c} 
\left( \frac{r-1}{r^4} \right) \e,
\end{equation} 
which again is small.  In fact,
even though this appears to be of the same order in $\e$ 
as the radiation  damping term, 
far from the black hole, it will be subleading since
 it contains a relative factor of $\sim \frac{1}{r}$,
whereas near the black hole, it actually becomes
$\frac{4a^2}{c} \e^2$.
This confirms our expectation
that the radiation damping term provides the dominant correction
to the equation of motion.

The $\e$-analysis verifies 
 that not only the expansion parameter, but also the ``self-field''
correction terms themselves in eq.(\ref{acorr}) may be taken 
arbitrarily small.
Hence, by letting $\e \rightarrow 0$, condition 3.\ is satisfied
in a self-consistent way, to desired accuracy (parameterized by $\e$).
However, 
this still is {\em not} a sufficient justification for neglecting 
the backreaction effects in our context, because
the difference between the equations of motion leading
to vastly different outcomes (i.e.\ particle falling in or bouncing)
is also tiny.
To see that explicitly, 
we now return to the first two conditions.
Although conditions 1.\ and 2.\ are of course automatically
guaranteed to hold in the test approximation by the construction process,
  it will be  useful
to determine the leading order in $\e$ at which this becomes evident.
This will illuminate the reason for the  condition 1.\ possibly
ceasing to hold in the first-order correction to the equation of motion.

First,  using eq.(\ref{rdsmin}), we have
\begin{equation} 
\rds(r_{m}) \sim 4 \left( \frac{a^2-b^2-c^2}{a^2-c^2} \right) \e^2
- 8 \left( \frac{ab}{a^2-c^2} \right) \e^3.
\end{equation} 
By the conditions on the coefficients $a,b,c$ imposed in
eqs.(\ref{qeps}),(\ref{Eeps}),(\ref{meps}), the coefficient of the
leading order in $\e$ is positive, so that for small $\e$,
$\rds(r_{m})>0$.
This means that according
to the test particle equation of motion, 
the particle falls past the horizon (and in fact, reaches the singularity), 
so that condition 1.\ is indeed satisfied in this approximation. 
We  also note that the particle overcharges the black hole:
\begin{equation} 
(M+E) -(Q+q) = - 2  (b-1)\, \e^2 <0
\end{equation} 
 by eq.(\ref{Eeps}).
This reconfirms condition 2.  

We may  now see why the test particle equation of motion
is not necessarily indicative of the actual motion, despite the fact
of the backreaction terms being small.
Whereas the particle ``slows down'' to $|\rdot(r_{m})| \sim \ord(\e)$,
the characteristic size of 
backreaction is also $ \ord (\e)$,
so that the
 backreaction effects may well ``destroy'' the process.
Furthermore,  we note that
the black hole is overcharged by only $ \ord (\e^{2})$,
though it does not seem meaningful to compare this estimate directly 
with the other quantities we discussed. 

We can also see 
how the backreaction could prevent the particle from 
being captured by the black hole
from a more physical standpoint:
As it falls towards the black hole, the particle loses energy to
electromagnetic and gravitational radiation.  
We may see this explicitly by evaluating
the change in energy along the particle's worldline:
for $u^{b} \, \nabla_{\! b}\, u^{a} = 
 \frac{q}{m} \, F^{ab} \,  u_{b} \, (1 + \stackrel{\sim}{h}\!(r))$,
\begin{equation} 
\label{dE}
u^{b} \, \nabla_{\! b}\, E 
= u^{b} \, \nabla_{\! b}\, \left[- \ddv (m \, u_{a} + q \, A_{a}) \right]
= -q \, \stackrel{\sim}{h}\!(r) \, f(r) \, \vdot(r)
\end{equation} 
which is negative, since both the radiation damping
and the (full) tail term give a positive contribution to 
$\stackrel{\sim}{h}\!(r)$, and $q$, $f$, and $\vdot$ are all positive.
This energy loss becomes stronger as the particle
approaches the horizon, so that near the horizon, 
the particle may no longer have enough energy to overcome the 
electrostatic repulsion of the black hole.  

To see then whether or not the process survives,  we would like to solve
the  corrected equation of motion, namely  
\begin{equation} 
\label{eomcorr}
  u^{b} \, \nabla_{\! b}\, u^{a} 
=  \frac{q}{m} \, F^{ab} \,  u_{b} 
+ \frac{2}{3} \, \frac{q^3}{m^2}  \,
   u^{c} \, \nabla_{\! c} \, (F^{ab}) \, u_{b} + \mbox{ ``tail''}
\end{equation} 
where again the local contribution to the ``tail'' is
$- \frac{q^2}{m} \,  u_{b} \, u_{c} \, \nabla^{[a}R^{b]c}$.
This is a rather complicated (integro-)differential 
equation, 
so solving it exactly
is intractable in the present situation.
However, working in the regime of interest where the correction 
terms are small,  we can achieve considerable simplification
by  using a linearized approximation.

The derivation of the test equation of motion, eq.(\ref{reom}),
depended only on the definition of the ``energy'' and the expression
for the 4-velocity $u^{a}$ of the particle, so the calculation is
identical in the present case, when $E$ is no longer a constant of 
motion.  Hence we can write:
\begin{equation} 
\label{reom_2}
\rds(r) =  
\frac{1}{m^2} \left[ E(r) - \frac{qQ}{r} \right]^2 - g(r).
\end{equation} 
We can find $E(r)$ by  solving
$\frac{dE}{dr} = \frac{1}{\rdot} \frac{dE}{d\tau} =
\frac{1}{\rdot} u^{b} \, \nabla_{\! b}\, E \approx
-q\, h(r) f(r) \vdot(r) / \rdot(r)$
(We have neglected the ``tail term''
contribution, since the above analysis indicated this 
to be subleading.)
Integrating, we obtain the energy as a function of $r$:
\begin{equation} 
\label{E(r)}
E(r) = E - q \, \int_{r}^{\infty} 
\frac{h(r') f(r') \vdot(r')}{|\rdot(r')|} dr' .
\end{equation} 
We now make the linearized approximation to $E(r)$ by letting
$h$, $\vdot$, and $\rdot$ in the integrand
be the expressions used in the test approximation
(i.e.\ dependent on the constant $E$ only).

Although the resulting expression is still too complicated to
determine its behavior analytically,
we may now find 
the corrected $\rds(r)$ numerically.
However, this constrains us to consider only specific examples,
so we cannot analyse the general behavior.
Such a calculation is discussed in section \ref{numex} below.
We find that in that case, $\rds(r)$ does become negative
at some $r_{0} > r_{+}$, which means that  the particle actually  ``bounces''
before reaching the horizon, so that the corrected process now
fails to destroy the horizon of the black hole.

\subsection{Numerical example}
\label{numex}

Here we discuss a specific example, obtained by following
the method outlined in section \ref{test}.
It will be convenient to 
choose our mass unit to correspond to the
(initial) black hole mass, i.e.\ we will set $M \equiv 1$.
(All other quantities are then treated as dimensionless.)

Consider a black hole with 
$Q = 0.99999 $, i.e.\ $M-Q = 10^{-5} $.  
Suppose the particle has charge 
$q = 3 \times 10^{-3} $,
mass $m = 1.8 \times 10^{-3} $,
and is thrown radially inward with energy
$E = 2.9897 \times 10^{-3} $.
In terms of the $\e$-expansion notation used above, we have: 
$\e \approx 0.0022$,
$a \approx 1.34$,
$b \approx 1.03$, and 
$c \approx 0.805$.
(As one may check explicitly, the conditions imposed
in eqs.\ (\ref{qeps})--(\ref{meps}) are all satisfied.)
We then have the following results:

First, the test particle equation of motion yields
$\rds > 0 \ \ \forall r$; in fact, 
$\rds(r_{m}) \approx 2.36 \times 10^{-6} $,
i.e.\ the particle is captured by the black hole.
(A full plot of $\rds(r)$ 
near its minimum is given in Fig.1.)

Second, the final total mass is $M_{\rm final} \leq M+E =
1.0029897$, 
whereas the final total charge
is $Q_{\rm final} = Q+q = 1.0029900$.  Thus, the
charge of the final configuration is greater
than the mass, $M_{\rm final} - Q_{\rm final} \leq -3 \times 10^{-7}$.
Although the process does not overcharge the black hole
significantly, this was to be expected from the fact that
we consider only ``nearly test'' processes.

Next, consider  the relative size of the  backreaction:  
Certainly all the parameters describing the
particle are much smaller than those corresponding
to the black hole: $q, E, m \ll M$.
Also, if we let the actual radius of the particle
 be, say, $r_{q} \approx 0.1 $, 
then the corresponding
stress energy tensor is small: 
$T_{\rm q} \approx 0.09 T_{\rm BH}$.
More importantly, the classical radius, 
(i.e.\ the expansion parameter for the self-field
corrections), is  tiny: 
$r_{\rm cl} = \frac{q^2}{m} \approx 0.005  \ll 1$. 
We may also check that the relative sizes of the
self-field terms are $\ll 1$;
the maximum attained by the radiation damping is 
$\sim h(r_{+}) \approx 0.0086$, while the 
local part of the tail term reaches the
maximal value of $\approx 0.001$.

Finally, we may compute the actual first order correction
to $\rds(r)$ as given by eqs. (\ref{reom_2}) and (\ref{E(r)}).
The corrected $\rds(r)$ is plotted in Fig.2. 
We see that now $\rds$ does become
negative, indicating that the particle 
does {\em not} fall past the horizon.

Since the first-order estimate of the backreaction effects
suggests that unlike the test case,
the particle actually bounces, 
one is led to ask what happens if we modify the 
parameters somewhat, in particular what if we increase the
particle's initial energy.
Since the upper bound on the energy was determined by the 
condition that the particle overcharges the black hole, it
suffices to require that only the energy the particle has
at the horizon, plus any additional energy that falls in from
the radiation, needs to satisfy the given bound.
Assuming that some radiation escapes to infinity, the starting
energy may then be higher.

To determine whether this relaxation is sufficient to allow
the particle to fall in, we consider the ``best case scenario,''
in which all the radiation escapes, so that the energy
which contributes to the final mass of the black hole is
just given by  the energy the particle has at the horizon.
Then we may use eq.(\ref{E(r)}) to determine what energy
the particle would have had at infinity.
Again, to avoid solving an integral equation, 
we use the approximation
\begin{equation} 
\label{E_inf(r)}
E_{\infty} = E(r_{+}) + q \, \int_{r_{+}}^{\infty} 
\frac{h(r') f(r') \vdot(r')}{|\rdot(r')|} dr'.
\end{equation} 
with the $E$ appearing in $\rdot$ and $\vdot$ being the (constant)
energy the particle has at the horizon, which we define
$E(r_{+}) \equiv Q + q - M$.
If the particle bounces even in this set-up, then 
backreaction cannot be overcome just by throwing in the
particle harder.

Using eq.(\ref{E_inf(r)}), we find that 
$E_{\infty} \approx 0.0122$, which is 
considerably larger than the final required energy. 
Our computations suggest that if this were indeed the starting
energy, then the particle would fall past the horizon, after all.
The corresponding plot of $\rds(r)$ is shown in Fig.3.
(However, given  the numerical accuracy required for this calculation,
one might question the sufficiency of the linearized 
approximation of the backreaction.  In particular, the minimum of 
$\rds \approx 6 \times 10^{-9} < \e^3$.)
Furthermore, if none of the escaped radiation contributed
to the mass of the black hole, the black hole would be 
(barely) overcharged.  However, since 
most of the lost energy is radiated near the horizon, such
a constraint does not seem too realistic.
Also, the first-order approximation which we used 
is not necessarily reliable near the horizon,
since $\frac{qhf\vdot}{|\rdot|} \sim \ord(\frac{\e^2}{(r-1)^2})$ 
which becomes $\sim \ord(1)$ at the horizon.

To summarize, we found that according to the test approximation, 
the particle will be captured by the black hole.
This would mean that the process overcharges the black hole,
leaving cosmic censorship in question.
The  first order correction to the equation of motion 
indicates that this will no longer hold, because the particle
will bounce.
However, since this conclusion depends crucially on the energy loss
near the horizon, where the linearized approximation may not be 
valid, and since our further calculation suggests that
increasing the initial energy of the particle may allow the particle
to fall past the horizon, we conclude that a more thorough analysis
would be required to reach a definitive answer.

\section{Imploding Charged Shell}
\label{shell}

In the previous section, we saw that even though the
backreaction effects may be kept arbitrarily small as compared
to the ``test'' case, the size of the process we are interested in
may be  smaller, so that the backreaction could easily prevent the
process from happening, or at least the final outcome 
is not readily analysable.

To avoid such complications arising from the backreaction, we are led 
to consider a simpler case in which these effects disappear,
namely the case of exact spherical symmetry \cite{kuchar_p}.
Spherically symmetric spacetimes have of course been studied
extensively in the past (for review of some of the studies
related to cosmic censorship, see e.g.~\cite{w2},\cite{sin}),
   the main appeal being that
in such cases, we may use the 
generalized Birkhoff's theorem and the appropriate matching conditions
to determine the dynamics exactly.  

The spherically symmetric analog of an infalling charged particle 
is an imploding charged shell.
Note that here we do not need to confine our
considerations to ``test'' shells.
Hence in this section, we  
consider the conditions for overcharging a black
hole with an imploding shell.
  
Pioneering efforts in this direction include
\cite{israel},\cite{kuchar},\cite{b}.
   In the special case  of an imploding charged
   shell in an otherwise empty spacetime, $Q=M=0$,
it was shown \cite{b} that the shell can not collapse to form 
a naked singularity.  
However, to the author's best knowledge, the more general
case of a shell imploding onto an already existing black hole
has not been fully analysed in the context of cosmic censorship.

Proceeding in parallel to section \ref{test}, we denote
the shell's total charge, rest mass, and energy by
$q$, $m$, and $E$, respectively.  The spacetime inside the
shell will be given by the Reissner-Nordstr\"{o}m metric, eq.(\ref{ds}),
for the black hole with 
$g(r)=g_{\rm in}(r) \equiv 1 - \frac{2M}{r} + \frac{Q^2}{r^2}$,
whereas outside the shell, the geometry will be described by
eq.(\ref{ds}) with $g(r)=
g_{\rm out}(r) \equiv 1 - \frac{2(M+E)}{r} + \frac{(Q+q)^2}{r^2}$,
which, for $Q+q > M+E$ corresponds to a ``naked singularity''
spacetime.
The radial coordinate is continuous across the shell; the position
of the shell will be denoted by $R$. 

The equation of motion for a charged shell is given by
\begin{equation} 
\label{s_eom}
\sqrt{g_{\rm in}(r) + \Rds} - \sqrt{g_{\rm out}(r) + \Rds} 
= -\frac{m}{r}
\end{equation} 
where $\dot{R} \equiv \frac{dR}{d\tau}$ and for an infalling shell we
can again write $\Rds$ as a function of $r$.
Solving for $\Rds(r)$, we obtain (cf.\ eq.(\ref{reom1}))
\begin{eqnarray} 
\label{s_reom1}
\Rds =  
 \left( \frac{E^2}{m^2} - 1 \right)
- \frac{2M}{r} \left[
\left( \frac{Q}{M} \frac{Eq}{m^2} - 1 \right)
+ \frac{E}{2M} \left( \frac{q^2}{m^2} - 1 \right) \right] 
\nonumber \\
+ \frac{Q^2}{r^2} \left( \frac{q^2}{m^2} - 1 \right)
\left[ 1 + \frac{q}{Q} + \frac{q^2-m^2}{4Q^2} \right].
\end{eqnarray} 
We can also write the equation of motion in a form
of eq.(\ref{reom}), but with an effective potential
$A_{v}^{\rm shell}(r) \equiv - \frac{Q}{r} - \frac{q}{2r} 
+ \frac{m}{q} \frac{m}{2r}$ 
which includes 
not only  the electromagnetic interaction with the black hole,
but also the electromagnetic 
self-repulsion and the gravitational self-attraction of the
shell.  (The factor of $2$ in the denominator of the latter two
terms comes from the standard ``Coulomb'' potential of a shell evaluated
on the shell itself.)  Hence 
\begin{equation} 
\label{s_reom}
\Rds =  
\frac{1}{m^2} \left[ E - \frac{qQ}{r} - \frac{q^2}{2r}  
+ \frac{m^2}{2r} \right]^2 - g_{\rm in}(r).
\end{equation} 

We now show that the conditions imposed on the parameters
are not mutually consistent.
As before, we require that the shell overcharges the black hole,
$M>Q$ and $M+E<Q+q$, and also that the parameters of the shell
satisfy $0<m<E<q$.
A shell which falls  past the horizon $r_{+}$ of the black hole
 has  $\Rds(r) > 0 \ \ \forall r\geq r_{+}$, which is the necessary
(and sufficient) condition for creating a naked singularity.
There are two physically distinct possibilities of how this
could be satisfied.

First, the shell may implode all the way down to the singularity,
which means $\Rds(r) > 0 \ \ \forall r$ and in particular,
\begin{equation} 
\label{s_Rdsmin}
\Rds(r_{m}) > 0
\end{equation} 
 where $r_{m}$ is the coordinate where $\Rds$
achieves its minimum.
Solving $\frac{d\Rds}{dr}(r_{m}) = 0$ for $r_{m}$ and 
writing $\Rds(r_{m})$ in 
a convenient form, we have:
\begin{equation} 
\label{s_Rdsmin1}
\Rds(r_{m}) =\frac{
- m^4 + 2 m^2 [\om - 2 E M -2 (M^2-Q^2)] - \om (\om - 4 E M)
- 4 E^2 Q^2 }{(q^2-m^2)[(q+2Q)^2-m^2]} 
\end{equation} 
where, for compactness of notation, we have defined
$\om \equiv q(q+2Q)$.
Since the denominator is always positive,
in order to satisfy eq.(\ref{s_Rdsmin}), we must find a suitable
combination of the parameters $(m,E,q,Q,M)$ such that the 
numerator is positive.
Imposing this as a condition on $m^2$, we must have
\begin{eqnarray} 
(\om - 2 E M -2 (M^2-Q^2))^2 -  \om (\om - 4 E M) - 4 E^2 Q^2  =
\nonumber \\
= 4 (M^2-Q^2) (E^2 + 2 E M + (M^2-Q^2) - \om) > 0.  
\end{eqnarray} 
However, this is exactly inconsistent with the ``overcharging'' 
conditions ($M>Q \ {\rm and} \  Q+q > M+E
\Rightarrow
 \om  > E^2 + 2 E M + (M^2-Q^2)$), which means that  eq.(\ref{s_Rdsmin})
can not be satisfied with any suitable parameters corresponding
to overcharging the black hole.
Thus, a naked singularity cannot result from a charged shell 
imploding to the central singularity.

The second, less stringent, way to create a naked singularity
is for the shell to implode past the horizon and ``bounce'',
i.e.\ $\Rds(r_{\rm bounce})=0$ with $r_{\rm bounce} < r_{+}$.
In this case, even though the shell does not itself crash into
the singularity, it does destroy the horizon of the black hole,
thereby exposing the singularity that already exists.\footnote{
It is not clear how this process might occur in a more realistic
black hole
collapse scenario, such as the black hole forming, say,  from a charged
ball of dust,  as opposed starting with an eternal black hole:  
There does not seem to be a consistent Penrose-Carter diagram
which would correspond to this process.}
Since $\Rds(r)$ has only one minimum, it is a monotonically
increasing function $\forall \ r>r_{\rm bounce}$, and in 
particular,
\begin{equation} 
\label{s_dRdsmin}
\frac{d\Rds}{dr}(r_{+}) > 0.
\end{equation} 
Eq.(\ref{s_dRdsmin}) is the analog of eq.(\ref{s_Rdsmin}),
and upon simplifications, we obtain the corresponding analog
of eq.(\ref{s_Rdsmin1}),
\begin{equation} 
\label{s_dRdsmin1}
\frac{d\Rds}{dr}(r_{+}) = \frac{
-  m^4 + 2 m^2 [\om -  E r_{+} -2 r_{+} \sqrt{M^2-Q^2}] 
- \om (\om - 2 E r_{+})}{2m^2 r_{+}^{3}}
\end{equation} 
which we require to be positive.

Now, the situation is a bit more complicated than in the previous case, 
because we {\em can} find some $m^2$ and $(E,q,Q,M)$ corresponding
to overcharging the black hole such that the RHS of 
eq.(\ref{s_dRdsmin1}) is positive.
However, the  physical constraints on $m^2$,
namely $0 < m^2 < E^2$, unfortunately turn out to be inconsistent
with these solutions. 
Although the analytic proof is not as apparent as above,
numerical calculations indicate that this is indeed the case.


This means that the shell must bounce before reaching the horizon,
so that it cannot create a naked singularity.
We note that in
the special case of $Q=M=0$, (i.e.\ a charged
shell in an otherwise empty spacetime) 
it is now easy to see that the shell cannot implode to form a
naked singularity.
In such  case 
eq.(\ref{s_reom1}) simplifies to 
\begin{equation} 
\label{s0_reom1}
\Rds =  
 \left( \frac{E^2}{m^2} - 1 \right)
- \frac{E}{r} \left( \frac{q^2}{m^2} - 1 \right)
+ \frac{m^2}{4r^2} \left( \frac{q^2}{m^2} - 1 \right)^{2},
\end{equation} 
which attains minimum at $r_{m} = \frac{q^2-m^2}{2E}$,
at which point $\Rds(r_{m}) = -1 < 0$.
Therefore, the shell bounces at $r_{\rm bounce} > r_{m} > 0$, and there
is no singularity anywhere.

What conclusions may one draw from this negative result?
There is an important difference between the shell and the
point particle: the shell experiences its own self-repulsion, whereas
the point particle does not.  On the other hand, the particle still
does experience the ``self-field'' effects from the
backreaction, which may prevent
it from falling past $r_{+}$.
We found that attempting to circumvent these backreaction
effects by considering the case of exact spherical symmetry
unfortunately failed to produce the desired effect.

\section{Extensions}
\label{ext}

The particular mechanism for possible violation of cosmic 
censorship discussed in section \ref{mech}
was motivated primarily by its algebraic simplicity.
However, we can imagine a number of possible generalizations
or extensions which could also lead to similar
results.

Following up on the lesson learned in the previous section, 
one would be tempted to try to find a ``compromise'' between
the shell and the particle, which would have the virtues of
both.  Namely, one would like to find a case in which
the backreaction effects would be small enough
to preserve the outcome of the process, and at the same time, 
the self-repulsion would not be too great to prevent the process
from happening in the first place.
One might hope that such a case could be achieved with a 
symmetric configuration of $N$ particles distributed uniformly
around the black hole and infalling  according to the same
equation of motion.
The previous two cases we discussed in sections \ref{mech} and
\ref{shell}
correspond to the special
cases of $N=1$ and $N \rightarrow \infty$, respectively.

Unfortunately, a definite conclusion about the 
backreaction  effects in such set-up 
 has not yet been reached.
Nevertheless, it is interesting to note some exploratory results
of such study.
We have considered explicitly the cases of 
$N=2$, $4$, and $6$ particles.
To find the electromagnetic force that each particle feels,
 one may use the
expression derived by~\cite{ll} for the potential of a point
charge in Reissner-Nordstr\"{o}m spacetime,
 along with the superposition principle.
One finds that according to the test process, one may still
find a finite (though more constrained) range of parameters
whereby all the particles fall past the horizon and overcharge
the black hole.  In this scenario, 
one is actually helped by causality:
Each particle ``feels'' the other ones as if
they were much further away from the black hole.
Indeed, one may find cases in which each ``process''
 by itself (i.e.\ each particle falling into the black hole
 in the field of the others)  is perfectly allowed
in all regimes (test as well as macroscopic) since by itself
this doesn't overcharge the black hole; only the combined
effect of all processes taken together leads to overcharging.
However, we expect that such configurations still do not reduce the
backreaction effects sufficiently, though 
more detailed calculations would be needed to verify it.

One may also imagine a number of generalizations
of the single-particle case which compensate for (or at least
alleviate) the
backreaction effects.
In the preceding treatment, we considered the particle moving
in a background spacetime with the only forces acting upon it
being determined by the background electrostatic field of the
black hole and the self-field of the particle.
However, one may also consider additional forces acting on the
particle.  In particular, by rescaling everything to a sufficient
size, one may imagine that the ``particle'' has a little rocket
attached to it, so that it may boost itself at any point of its
trajectory.
Then the particle could possibly ``compensate'' for the backreaction
effects, by boosting itself to the appropriate energy just before
crossing the horizon, 
so that it would fall in, 
while still overcharging
the black hole.

Perhaps the most interesting extension of the
basic mechanism is to consider angular momentum
as well as charge as free parameters.
For example, we could send the particle(s) toward 
the horizon spinning, or in a non-radial direction,
such that it (they) would impart angular momentum to 
the black hole.  The corresponding equation of motion would 
of course be far more complicated, but from a physical
standpoint, such cases are more realistic and perhaps
more ``promising'', since extremality might possibly be
supersaturated more easily:   If the black hole 
captures the particle(s), both the charge and the
angular momentum of the black hole would increase.
(Again, we stress that it is crucial to start
with a near extremal black hole.)

An interesting limit to consider is the completely
uncharged case, namely throwing a spinning particle
into a near extremal Kerr black hole.  
This case has several appealing features.
First, we do not have to worry about the 
electromagnetic self-field  backreaction
effects, which in the original mechanism were the
most constraining and nontrivial effects.
Second, this would present a much more physically realistic 
case, since we believe that near-extremal
Kerr black holes actually exist in our universe.
Finally, the causal structure inside the horizon
of a Kerr black hole is even more interesting than
that inside a Reissner-Nordstr\"{o}m black hole.

Although we have not examined these latter extensions,
which would probably be better addressed by means of numerical
simulations,
it seems that they might yield a more definitive answer as
to whether cosmic censorship is upheld or not.

\section{Discussion and Conclusions}
\label{concl}

We have reexamined the old question of
cosmic censorship in the
simple context of a charged particle falling into a
near extremal Reissner-Nordstr\"{o}m  black hole.
In particular, we tried to analyse
whether a particle which would overcharge the black hole
could fall past the horizon.
As in earlier works which considered this type of set-up
(i.e.\ particles supersaturating  extremal bounds 
of  black holes),
such as \cite{w1},\cite{co},\cite{n},\cite{semiz},\cite{br},
the basic motivation was that if the particle 
 is actually captured by the black hole,
then it seems that the resulting final configuration
cannot sustain an event horizon,\footnote{
Technically, in such a situation, there never existed an
event horizon, but only the apparent horizon.}  
suggesting that 
a naked singularity results.
This conclusion appears even more inescapable if the particle
falls all the way into the singularity.\footnote{ 
Then one doesn't need to worry about the scenario
where the particle reemerges, possibly leaving behind 
a near extremal black hole again.}

The simplest way to examine the possibility of
supersaturating the extremal bound of a black hole
is to start at extremality and consider strictly test processes.
Unfortunately,  such processes cannot lead to destroying the horizon, since
the particles which are allowed to fall in would at best 
maintain extremality
 (\cite{w1},\cite{co},\cite{n},\cite{semiz}).
Faced with these results, one is led to 
start with a near extremal black hole.
One might then hope that such a set-up allows the black hole
to ``jump over'' 
extremality, by capturing a particle with
appropriate parameters.

Perhaps somewhat surprisingly, 
we found that according to the test particle approximation,
this can be indeed achieved.
Namely, a broad range of configurations can be found in which,
according to the test equation of motion, 
the particle falls into the singularity and the resulting
object exceeds the extremal bound.
This result, which seems to have been previously overlooked,
is important in the context of these earlier attempts to violate
cosmic censorship.

However, we now have to contend with backreaction.
The natural question then arises, ``how good is the
test approximation in this context?''
We have seen that this issue is somewhat subtle, 
and in fact it is not yet fully resolved.
In particular, the backreaction effects
may be kept  arbitrarily small as compared to the leading order
effects, but this does not suffice.
Indeed, we have seen from our numerical example that
the particle may lose enough energy due to the backreaction
effects that it bounces back to infinity instead of being
captured by the black hole.

Although we have not found good examples in which the particle
falls past the horizon even when the first-order effects
of the backreaction are taken into account, we have not 
found a solid indication to the contrary.
Furthermore, it is not clear 
 that the situation described by
our example is in fact generic in terms of the outcome.
Also, the possible extensions mentioned in the previous
section, which may lead to the particle falling past the
horizon after all,
have not been analysed.

On the other hand,
 even if the particle is ``captured'' by the black hole,
the resulting implications for cosmic censorship are
no longer as clear as in the test case.
We have seen that the backreaction effects are important
already  before the particle crosses the horizon; 
they would become dominant after.
This means that inside the horizon, we definitely can not
trust the test particle equation of motion.
It way well happen that the self-field effects will cause
the particle to ``bounce'' at some radius $r_{0} < r_{+}$,
or even that the language we have been using to describe
the situation is no longer appropriate.
In the former case, the particle reemerges (possibly into
the same asymptotically flat region it started from, since
there can not have been any horizon ``above'' the particle),
and the inner singularity may never become visible to the
outside observer.

Given all these considerations, we are led to conclude that
even in the case of a single particle falling into a black hole, 
a more detailed analysis will be needed to settle the issue
of whether or not the cosmic censorship prevails.

\bigskip\bigskip
\noindent{\large \bf Acknowledgements}
\medskip

It is a great pleasure and honor
to thank Gary Horowitz
for many useful discussions, for reading the manuscript, and
for all his guidance and support.
I also wish to thank
Sean Carroll, Doug Eardley,
Karel Kucha\v{r}, Simon Ross, Harrison Sheinblatt,
Bob Wald, and Haisong Yang
for helpful discussions,
and Jim Hartle and Bei-Lok Hu for initial encouragement.
I acknowledge UCSB Doctoral Scholar Fellowship
and DoD NDSEG  Fellowship  for financial support.

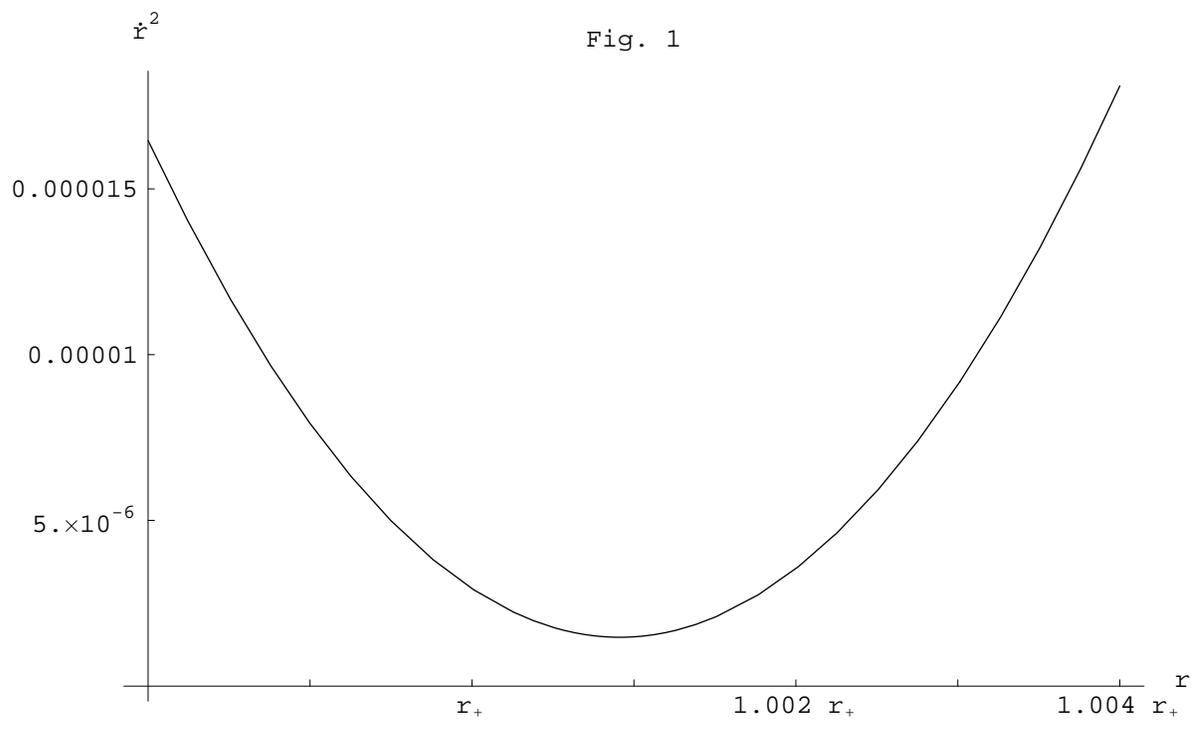
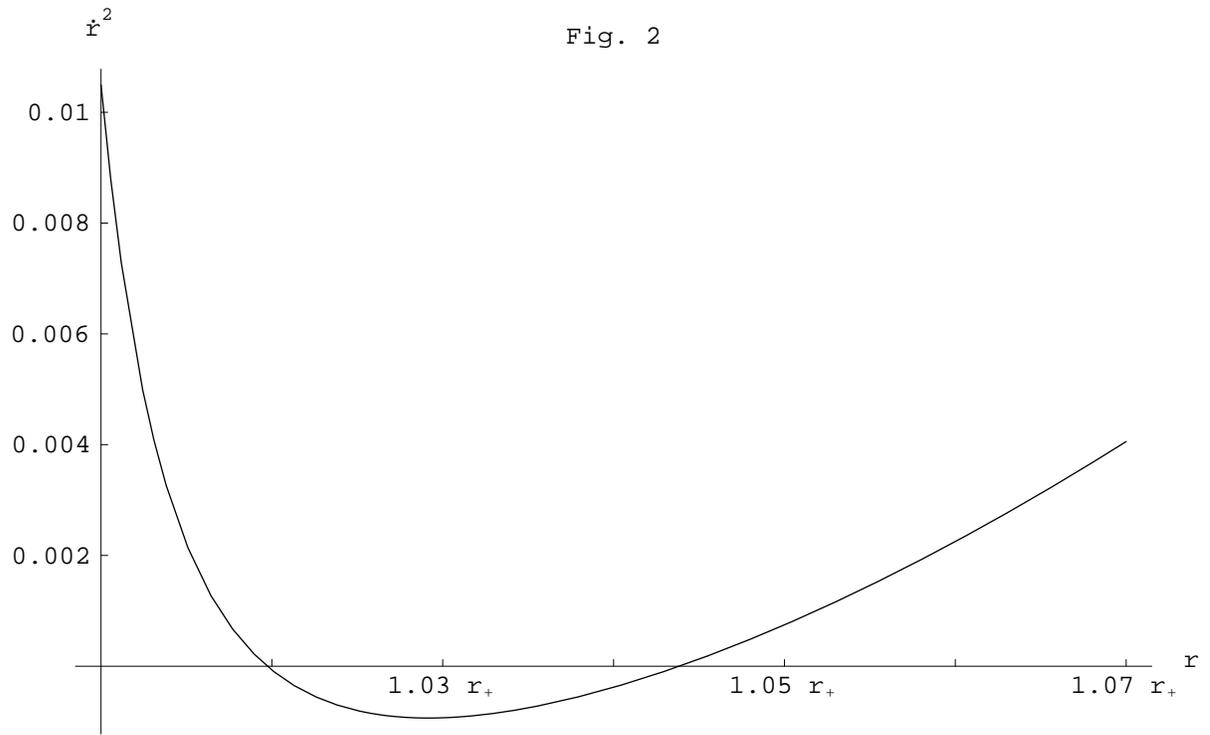
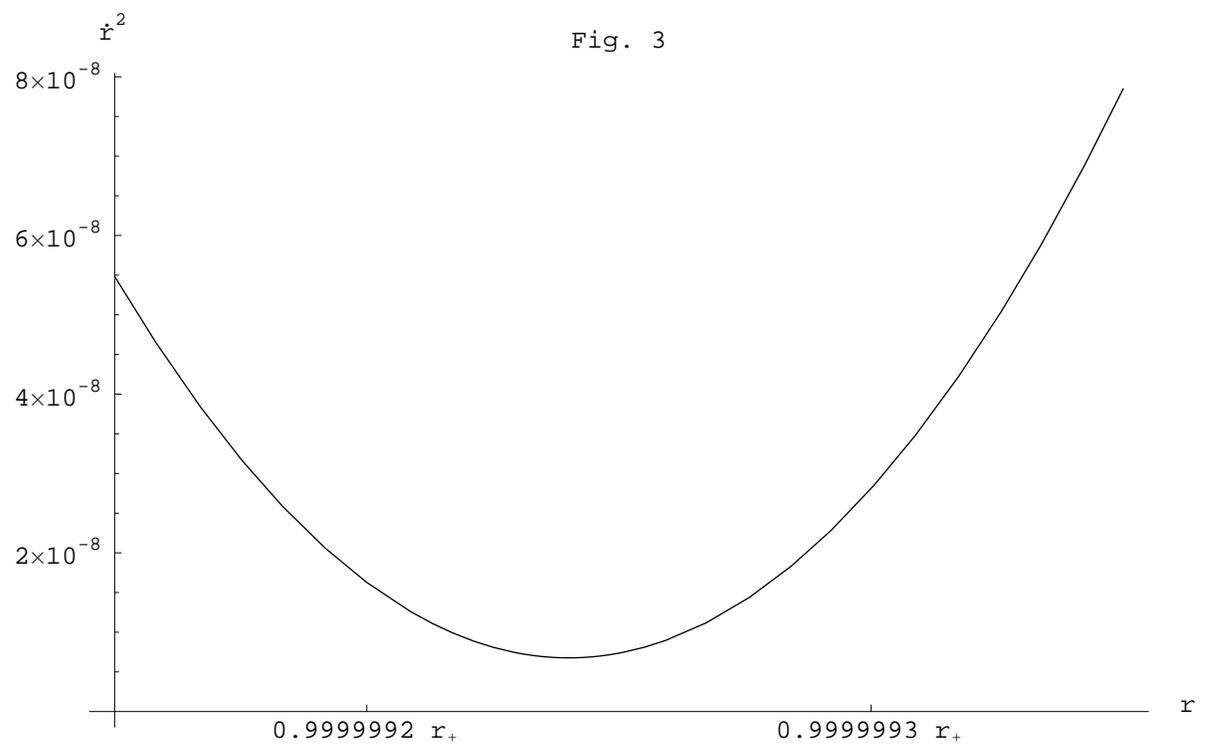
\end{document}